\documentclass[%
 reprint,
 amsmath,amssymb,
 prb,
 floatfix,
]{revtex4-2}

\usepackage{graphicx}
\usepackage{dcolumn}
\usepackage{bm}
\usepackage{mathtools}
\usepackage{bm}
\usepackage{graphicx}
\usepackage{dcolumn}
\usepackage{bm}
\usepackage{epstopdf}
\usepackage{xcolor}
\usepackage{xspace}
\usepackage[T1]{fontenc}
\usepackage{lmodern} 

\usepackage{booktabs}
\usepackage{hyperref}
\usepackage[all]{hypcap}
\hypersetup{colorlinks, linkcolor=blue,          
    citecolor=blue,        
    filecolor=blue,      
    urlcolor=blue           
}

\begin{document}

\preprint{APS/123-QED}

\title{Field evolution of the magnetic structure and spin Hamiltonian in Cs$_2$RuO$_4$}

\author{S.D. Nabi$^1$}
 \email{nabid@ethz.ch}
\author{E. Ressouche$^2$}%
\author{D. G. Mazzone$^3$}%
\author{J. Lass$^3$}%
\author{R. Sibille$^3$}%
\author{Z. Yan$^1$}%
\author{S. Gvasaliya$^1$}%
\author{A. Zheludev$^1$}%
 \email{http://www.neutron.ethz.ch/}
\affiliation{%
 $^1$Laboratory for Solid State Physics, ETH Zurich, 8093 Z\"urich, Switzerland\\
 $^2$Universit\'e Grenoble Alpes, CEA, IRIG, MEM, MDN, 38000 Grenoble, France\\
 $^3$PSI Center for Neutron and Muon Sciences, Paul Scherrer Institute, 5232 Villigen, Switzerland}%
 
\date{\today}

\begin{abstract}

We report neutron diffraction under applied magnetic fields and complementary zero-field neutron spectroscopy measurements on Cs$_2$RuO$_4$. Previous work [Phys. Rev. B. 112, 134436 (2025)] identified a spin-flop-like transition accompanied by a quantum critical point within the ordered phase, attributed to strong frustration between alternating single-ion anisotropy planes. Here, we quantitatively confirm the predicted field evolution of the magnetic structure using neutron diffraction. Furthermore, analysis of the excitation spectrum within an SU(3) spin-wave framework resolves previously undetermined parameters of the minimal spin Hamiltonian and lifts the associated degeneracies.

\end{abstract}

\maketitle


\section{\label{sec:level1}Introduction\protect\\}

Cs$_2$RuO$_4$ belongs to the well-studied family of frustrated quantum magnets Cs$_2MX_4$, whose members realize a diverse range of effective exchange geometries, giving rise to a wide range of exotic magnetic phenomena. These include the distorted triangular lattice realized in Cs$_2$CuCl$_4$ \cite{CCC1, CCC2, CCC3, CCC4} and Cs$_2$CuBr$_4$ \cite{CCB1, CCB2}, the effectively one-dimensional chain in Cs$_2$CoCl$_4$ \cite{CCoC1, CCoC2, CCoC3}, and the frustrated zigzag ladder structures found in Cs$_2$CoBr$_4$ \cite{Povarov2020, Facheris2022, Facheris2023, Facheris2024} and Cs$_2$CoI$_4$ \cite{CCI}.

\begin{figure}[t]
    \centering
\includegraphics[clip, trim=0.0cm 0.0cm 0.0cm 0.0cm ,width=0.7\linewidth]{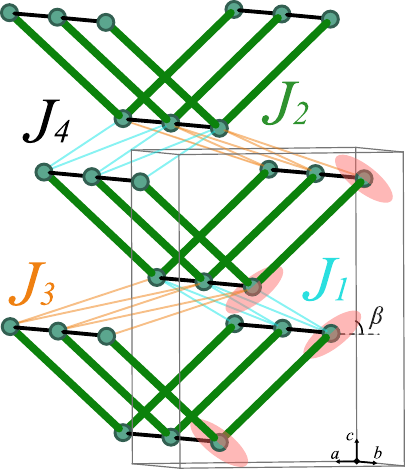}
\caption{\label{fig:exchanges} Schematic overview of interaction network in Cs$_2$RuO$_4$. $J_8$ is left out for visibility. The anisotropy planes (rose) are only shown within the drawn unit cell (gray).}
\end{figure}

The title compound has recently been identified as a unique member of this family \cite{CRO}. Substituting the 3$d$ transition metal with a 4$d$ Ru$^{6+}$ ion ($S = 1$), and replacing the halogen ligand with oxygen, leads to a profound modification of the exchange interaction hierarchy. In contrast to other members of the family, the exchange topology in Cs$_2$RuO$_4$ is strongly three-dimensional, as discussed in detail in \cite{CRO} and schematically illustrated in Fig.~\ref{fig:exchanges}. The dominant exchange interaction, $J_2$, couples ions whose single-ion anisotropy planes ($D$) are nearly orthogonal and alternate with a tilt angle $\beta$ relative to the \textbf{a} axis. This geometry produces a pronounced frustration of the anisotropy planes, exceeding that observed in previously studied compounds of the family.

In zero magnetic field (ZF), the system adopts a stripe N\'eel-ordered ground state below $T_\mathrm{N}$ = 5~K \cite{CRO}, with spins aligned along the effective easy axis \textbf{b}, defined by the intersection of the anisotropy planes. The magnetic field–temperature ($H$–$T$) phase diagram reveals a field-induced continuous spin-flop-like transition accompanied by a quantum critical point (QCP) within the ordered phase. A minimal chain model, capturing the dominant $J_2$ interaction between alternating anisotropy planes, reproduces this behavior and attributes the QCP to the strong frustration inherent to the system. Notably, this transition is characterized by the spontaneous emergence of a transverse staggered magnetization coexisting with the longitudinal component.

Despite this progress, several key aspects remain unresolved. In particular, the field evolution of the magnetic structure has not yet been directly determined. Furthermore, the minimal model includes additional exchange parameters ($J_1$ and $J_3$, shown in Fig.~\ref{fig:exchanges}) that could not be independently extracted from previous spectroscopic measurements. The tilt angle $\beta$ was also determined to be $45^\circ$, leaving the orientation of the high-field ground state ambiguous between \textbf{a} and \textbf{c}.

In this work, these open questions are addressed using neutron diffraction measurements under applied magnetic fields, directly confirming the predicted evolution of the magnetic structure across the QCP. Complementary ZF neutron spectroscopy measurements, performed in an orthogonal sample orientation relative to \cite{CRO}, lift the remaining degeneracies, enabling independent resolution of the previously undetermined Hamiltonian parameters and establishing a refined minimal model for Cs$_2$RuO$_4$.

\section{\label{sec:level1}Methods\protect\\}

Cs$_2$RuO$_4$ shares the orthorhombic $Pnma$ (No. 62) structure common to the Cs$_2MX_4$ family, with room-temperature lattice constants $a$ = 8.528(2), $b$ = 6.494(2), and $c$ = 11.498(2)~\AA. The dark green single crystals used in this study were obtained by self-flux growth in a platinum crucible, and their structural quality was confirmed by single-crystal x-ray diffraction on a Bruker APEX-II diffractometer; full refinement details are reported in \cite{CRO}.

Neutron diffraction experiments were conducted on the lifting counter diffractometers D23 at the Institut Laue Langevin (ILL) and Zebra at the Paul Scherrer Institute (PSI). In both experiments the same 130~mg sample was mounted with the $(h~0~l)$ plane in the scattering plane of the instrument. The measurements were performed in conjunction with a vertical 6~T cryomagnet, with the applied field along the \textbf{b} axis at a base temperature of 1.5~K. At D23, a wavelength of $\lambda$ = 1.28~\AA~was selected using a Cu(200) monochromator and at Zebra, $\lambda$ = 1.18~\AA~using a Ge(113) monochromator. Nuclear refinements were carried out using the \texttt{FullProf Suite} software \cite{Fullprof}.

Neutron spectroscopy experiments were performed on the same 1~g sample of Cs$_2$RuO$_4$ used in \cite{CRO}, at the multiplexing spectrometer CAMEA \cite{CAMEARSI} at PSI. The sample was installed with the ($h$~$0$~$l$) plane horizontal in the scattering plane of the spectrometer, and mounted in a cryomagnet. Foreground measurements were collected at a base temperature of $T$ = 1.6~K. Four measurement series
were performed using incoming neutron energies of $E_i$ = 4.87, 4.92, 4.96, and 5~meV (elastic resolution $\sim$~0.19~meV). Sequential steps of approximately 0.08 meV were used to interlace the datasets and suppress normalization artifacts. Scattering angles of 2$\theta$ = -41$^{\circ}$, -45$^{\circ}$ were used. For each 2$\theta$, the sample was rotated over a 110$^{\circ}$ range in 1$^{\circ}$ steps, counting 60~s per angle. Background data were collected with 1/3 of the counting statistics at 50,
100, and 150~K over the same range. Data reduction and background subtraction were performed using the \texttt{MJOLNIR} \cite{MJOLNIR} software package. SU(3) spin wave theory calculations were performed using \texttt{SUNNY.jl} \cite{sunny}.

\section{\label{sec:RES}Results and data analysis\protect\\}

\subsection{\label{sec:DIFF}Neutron diffraction}

Neutron diffraction experiments were performed to determine the field evolution of the magnetic propagation vector and magnetic structure across the QCP. The ZF structure was previously established as a stripe N\'eel state with spins along the effective easy axis \textbf{b} \cite{CRO}, as shown in Fig.~\ref{fig:magstruc}(a). 

\subsubsection{Domain treatment}

Due to a high-temperature hexagonal-to-orthorhombic structural phase transition during crystal growth \cite{structrans}, the sample contains three crystallographic domains sharing the \textbf{a} axis and rotated by approximately $60^{\circ}$ with respect to each other. Since $|\textbf{b}|/|\textbf{c}| \approx \sqrt{3}$, the structure remains close to hexagonal symmetry. As a result, a $\pm 60^{\circ}$ rotation maps any $(h~k~l)$ reflection onto two additional positions and vice versa.

The position of these mapped reflections depends on the parity of $k+l$: for even $k+l$, they fall on integer indices, while for odd $k+l$, they appear at half-integer positions. For example, the (0~0~2) reflection overlaps with (0~1~1) and (0~$-1$~1), whereas (1~0~1) overlaps with (1~0.5~0.5) and (1~0.5~$-0.5$).

For nuclear Bragg peaks, this leads to a parity-dependent domain contribution: when $k+l$ is even, all three domains contribute, whereas for odd $k+l$, only the main domain contributes in the absence of magnetic order ($T \gg T_\mathrm{N}$). In contrast, magnetic Bragg peaks associated with the ZF propagation vector $\bm{k} = (0~0.5~0.5)$ occur at half-integer indices. In this case, the measured intensity contains one nuclear contribution from a rotated domain—which can be removed by background subtraction at $T \gg T_\mathrm{N}$, and two magnetic contributions from the main domain and one rotated domain.

To account for domain overlap, two approximations are adopted: (i) the relative rotations between domains are taken to be exactly $\pm 60^{\circ}$, and (ii) overlapping reflections are assumed to occur at identical $2\theta$. The latter approximation is valid at small scattering angles; at larger angles, deviations from ideal hexagonal symmetry lead to a gradual splitting of the peaks.

\begin{figure}[t]
\centering
\includegraphics[clip, trim=0.0cm 0.0cm 0.0cm 0.0cm ,width=\linewidth]{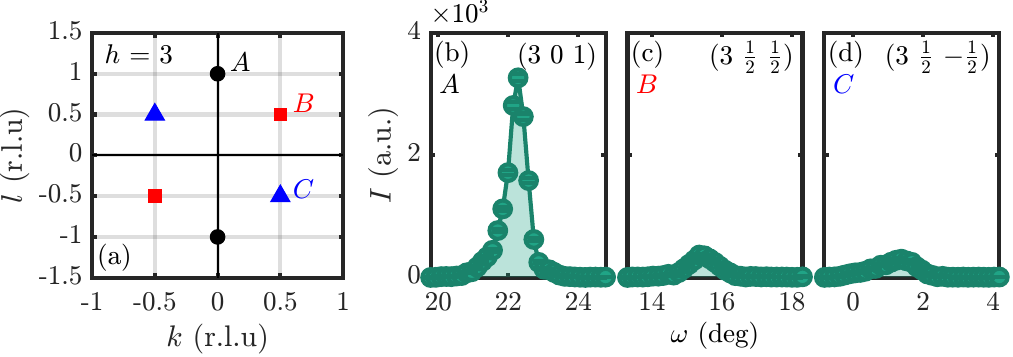}
\caption{\label{fig:massratiodiff} (a) (3~$k$~$l$) plane illustrating the (3~0~$\pm$1) peaks of the correctly aligned domain $A$ (dots), and the rotated domains $B$ (squares) and $C$ (triangles). (b-d) show rocking curves on a nonoverlapping (3~0~1) peak of domains $A$, $B$, and $C$, respectively. The ($h$~$k$~$l$) position on the top right indicates the indexation based on the alignment of crystallographic domain $A$.}
\end{figure}

Within this framework, a rocking curve measured at a Bragg position $(h~k~l)$ of the aligned domain $A$ contains contributions from $\mathfrak{R}{-}{(h~k~l)}$ of domain $B$ and $\mathfrak{R}{+}(h~k~l)$ of domain $C$, where $\mathfrak{R}{\pm}$ denotes a $\pm 60^{\circ}$ rotation. The measured integrated intensity can therefore be written as

\begin{equation}
I_{(h~k~l)} = A I_{(h~k~l)} + B I_{\mathfrak{R_-}(h~k~l)} + C I_{\mathfrak{R_+}(h~k~l)},
\label{eq:summassrationucdom}
\end{equation}

where $A$, $B$, and $C$ are the domain population ratios.

These ratios are determined independently from a nuclear Bragg peak whose rotated counterparts fall at half-integer $k$ and $l$, ensuring separation of all three contributions in the absence of magnetic order. The (3~0~1) reflection satisfies this condition, mapping onto (3~0.5~0.5) and (3~$-0.5$~0.5) in the reference frame of domain $A$, measured at 15~K ($T > T_\mathrm{N}$). This is illustrated in Fig.~\ref{fig:massratiodiff}(a). From the integrated intensities of these peaks [Fig.~\ref{fig:massratiodiff}(b–d)], the domain population ratios are obtained as

\begin{equation}
A = 0.762(3), \quad B = 0.123(3), \quad C = 0.115(3).
\label{eq:massrationucdom}
\end{equation}

The sample was aligned such that the $(h~0~l)$ plane of domain $A$ lies in the scattering plane, corresponding to a magnetic field applied along \textbf{b}. For domains $B$ and $C$, the field is rotated by approximately $\pm 60^{\circ}$ within the \textbf{bc} plane, i.e., close to the [0~$\pm$1~3] directions, which are symmetry-equivalent in the orthorhombic structure.

In ZF, all domains share the same magnetic structure, and their contributions can be treated straightforwardly. In applied field, however, the domains respond differently. Based on the minimal model Hamiltonian of Ref.~\cite{CRO}, a field along [0~$\pm$1~3] does not induce a phase transition, but instead produces a uniform moment and a canting of the staggered moment toward \textbf{c}. Accordingly, domains $B$ and $C$ are approximated as remaining in the ZF N\'eel stripe state. This approximation is supported by the refinement results presented below.

\begin{figure}[t]
\centering
\includegraphics[clip, trim=0.0cm 0.0cm 0.0cm 0.0cm ,width=0.6\linewidth]{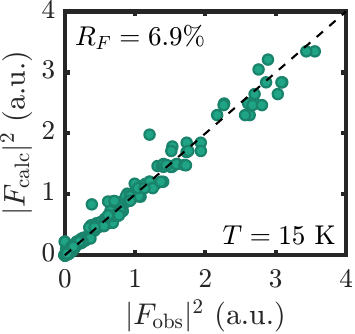}
\caption{\label{fig:nucref} Calculated vs observed integrated intensities of nuclear Bragg peaks at $T$ = 15~K.}
\end{figure}

\subsubsection{Nuclear refinement}

Now that the domain treatment and mass ratios are established, the nuclear structure was refined at 15~K, above $T_\mathrm{N}$, to determine the atomic positions and the overall scale factor required for subsequent magnetic refinements. To this end, 315 nuclear Bragg peaks were collected as rocking curves counting 1~s per point.

A baseline background was fitted and subtracted from each scan, after which the intensities were obtained by numerical integration and corrected for the Lorentz factor. The resulting intensities were modeled, where appropriate, as a sum of up to three nuclear structure factor contributions from the different domains according to Eq.~\ref{eq:summassrationucdom}, with the number of contributing domains determined by the parity of $k+l$.

The refinement included 20 parameters in total: 13 atomic positions, six isotropic thermal displacement parameters, and one overall scale factor. The refined atomic positions are consistent with the room-temperature structure, while the thermal displacement parameters are reduced, as expected. Including extinction parameters did not improve the fit.

The overall agreement factor is $R_F$ = 6.9\%, with calculated versus observed intensities shown in Fig.~\ref{fig:nucref}. The refined Ru$^{6+}$ positions, which are used in the magnetic structure determination, are listed in Table~\ref{tab:rupositions}.


\begin{figure*}[t]
\centering
\includegraphics[clip, trim=0.0cm 0.0cm 0.0cm 0.0cm ,width=\linewidth]{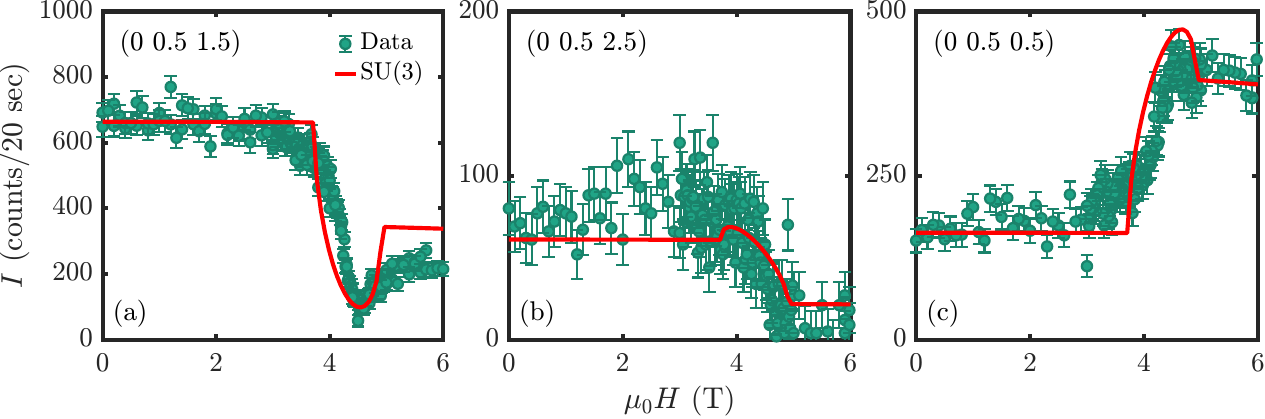}
\caption{\label{fig:fieldscanop} Background subtracted order parameter scans as a function of applied magnetic field along the \textbf{b} direction on magnetic Bragg peaks (a) (0~0.5~1.5), (b) (0~0.5~2.5), and (c) (0~0.5~0.5). Red lines indicate simulated scans using SU(3) spin wave theory. Peak (a) was measured on both instruments and scaled to match. (b-c) were only measured on D23.}
\end{figure*}

\begin{table}[t]
    \centering
    \caption{Refined Ru$^{6+}$ fractional coordinates at 15~K.}
    \label{tab:rupositions}
    \begin{tabular*}{\columnwidth}{@{\extracolsep{\fill}}lccc}
        \toprule
        Ion & $x$/|\textbf{a}| & $y$/|\textbf{b}| & $z$/|\textbf{c}| \\
        \midrule
        Ru$_1$ & 0.274(1) & 0.75 & 0.421(1) \\
        Ru$_2$ & 0.774(1) & 0.75 & 0.078(1) \\
        Ru$_3$ & 0.726(1) & 0.25 & 0.578(1) \\
        Ru$_4$ & 0.226(1) & 0.25 & 0.922(1) \\
        \bottomrule
    \end{tabular*}
\end{table}

\subsubsection{Order parameter evolution}

The ZF magnetic propagation vector \textbf{\textit{k}} = (0~0.5~0.5) \cite{CRO} was tracked as a function of applied field up to 6~T by monitoring the intensities of three magnetic Bragg peaks on D23 and Zebra, with counting times of 20 and 90~s per field, respectively. A baseline background was subtracted using intensities measured at 15~K, above $T_\mathrm{N}$. The resulting field dependences are shown in Fig.~\ref{fig:fieldscanop}.

The three reflections, (0~0.5~1.5), (0~0.5~2.5), and (0~0.5~0.5), were selected based on their domain-overlap properties. For each reflection, one of the two domain-related overlaps coincides with a symmetry-forbidden nuclear position, (0~1~0), (0~$-1$~2), and (0~0~1), respectively, thereby minimizing nuclear background and excluding any ferromagnetic contribution in applied field. In addition, (0~0.5~1.5) and (0~0.5~0.5) overlap with equivalent reflections from a second domain, whereas (0~0.5~2.5) overlaps with (0~1.5~0.5), which carries negligible intensity in ZF, and at 6~T all the intensity in Fig.~\ref{fig:fieldscanop}(b) is essentially in the background. Under the approximation that the minority domains remain close to the ZF N\'{e}el state, their contributions provide only a field-independent offset, such that the observed field evolution is dominated by domain $A$.

As shown in Fig.~\ref{fig:fieldscanop}, the propagation vector (0~0.5~0.5) is preserved throughout the entire field range. No additional propagation vectors are observed at high field, apart from the ferromagnetic component at (0~0~0) inferred from the appearance of bulk magnetization.

For all three reflections, the intensity remains approximately constant up to 4~T, consistent with the system remaining in the ZF N\'{e}el state. This further supports the assumption that, at least up to this field, the misoriented domains do not undergo significant changes. Above 4~T, across the QCP \cite{CRO}, a clear continuous evolution is observed: the intensity of (0~0.5~1.5) decreases sharply, nearly vanishing around 4.5~T before partially recovering and saturating toward 6~T. The (0~0.5~2.5) reflection decreases more gradually and becomes indistinguishable from the background above 4.5~T. In contrast, (0~0.5~0.5) shows the opposite trend to (0~0.5~1.5), reaching a maximum near 4.5~T.

The polarization-factor dependence of these reflections is consistent with a continuous reorientation of the staggered moment from \textbf{b} toward \textbf{c}, as confirmed quantitatively by the magnetic structure refinements presented below.

\subsubsection{Magnetic structure refinement framework}

Magnetic structures are described within the propagation-vector formalism, following the procedure outlined in Ref.~\cite{Facheris2024}.

The magnetic moment on ion $j$ in unit cell $n$ at position $\bm{R}_n$ can be expressed as a Fourier expansion,
\begin{equation}
\bm{m}_{n,j} = \sum_{\bm{k}}
\bm{A}_j^{(\bm{k})}
e^{-i\bm{k}\cdot\bm{R}_n},
\label{eq:momentfourier}
\end{equation}

where $\bm{k}$ is the magnetic propagation vector defining the periodicity of the moment arrangement relative to the crystallographic unit cell. The magnetic structure is thus parametrized by $3\times4$ complex Fourier components $\bm{A}_j^{(\bm{k})}$. Symmetry constrains these components to linear combinations of basis vectors $\psi_\nu$ belonging to irreducible representations (irreps) of the little group, i.e., the subgroup of the space group that leaves $\bm{k}$ invariant. The coefficients $C_\nu$ of these basis vectors are refined to reproduce the measured Bragg intensities.

In the presence of an applied magnetic field, a uniform moment with propagation vector (0~0~0) is induced on each ion. The total magnetic moment can therefore be written as $\bm{m}_{n,j} = \bm{m}^{(0~0.5~0.5)}_{n,j} + \bm{m}^{(0~0~0)}_{n,j}$, introducing an additional $3\times4$ parameters associated with $\bm{A}_j^{(0~0~0)}$. However, the corresponding Bragg peaks coincide with much stronger nuclear reflections and cannot be reliably separated in unpolarized neutron diffraction. The (0~0~0) component is thus only indirectly accessible through the total magnetization. In the refinements below, the uniform \textbf{b} component is constrained using bulk magnetization data from Ref.~\cite{CRO}, assuming equal contributions from all ions in the unit cell. Within the minimal model of Ref.~\cite{CRO}, transverse uniform components are also expected to emerge above the QCP, however, these cannot be directly accessed experimentally and are therefore left unconstrained.

For propagation vector (0~0.5~0.5) in space group $Pnma$, and using the atomic positions listed in Table~\ref{tab:rupositions}, only two two-dimensional irreps are symmetry-allowed. The six basis vectors $\psi_\nu$ of the irrep corresponding to the ZF structure ($\Gamma_2$) are listed in Table~\ref{tab:basisvectors}.

\begin{table*}[t]
    \centering
    \caption{Basis vectors associated with the observed irreducible representation $\Gamma_2$ in ZF for each of the magnetic atoms in the crystallographic unit cell.}
    \label{tab:basisvectors}
    \begin{tabular*}{\textwidth}{@{\extracolsep{\fill}}lcccccc}
        \toprule
        $\Gamma_2$ & $\psi_1$ & $\psi_2$ & $\psi_3$ & $\psi_4$ & $\psi_5$ & $\psi_6$ \\
        \midrule
        Ru$_1$ & (1,0,0) & (0,1,0) & (0,0,1) & (1,0,0) & (0,$-$1,0) & (0,0,1) \\
        Ru$_2$ & (1,0,0) & (0,$-$1,0) & (0,0,$-$1) & (1,0,0) & (0,1,0) & (0,0,$-$1) \\
        Ru$_3$ & (1,0,0) & (0,1,0) & (0,0,1) & ($-$1,0,0) & (0,1,0) & (0,0,$-$1) \\
        Ru$_4$ & (1,0,0) & (0,$-$1,0) & (0,0,$-$1) & ($-$1,0,0) & (0,$-$1,0) & (0,0,1) \\
        \bottomrule
    \end{tabular*}
\end{table*}

The $\Gamma_2$ irrep supports two magnetic domains, related by the generators of the little group:
\begin{align}
\mathrm{GEN}_1 &= -x, -y, -z \nonumber \\
\mathrm{GEN}_2 &= -x+1/2, -y, z+1/2 \nonumber \\
\mathrm{GEN}_3 &= -x, y+1/2, -z.
\end{align}
GEN$_1$ inverts all spins by $180^\circ$ and leaves the structure factors unchanged. GEN$_2$ and GEN$_3$ interchange $\psi_{1,2,3}$ with $\psi_{4,5,6}$, producing distinct structure factors for the two domains. As the basis vectors are symmetry-related, the same coefficients $C_\nu$ apply to both domains. Since the sample was ZF cooled and the magnetic field was applied within the ordered phase, no preferential domain population is assumed a priori. Equal domain populations are used as a starting point, with deviations tested.

The structural domain ratios $A$, $B$, and $C$ determined in Eq.~\ref{eq:massrationucdom} are included in all refinements. Misaligned domains are assumed to retain the ZF structure, and for each reflection the overlapping contributions from rotated domains are properly indexed. The polarization factor is evaluated at the $\bm{q}$ position defined in the reference frame of the aligned domain $A$.

The magnetic form factor for Ru$^{6+}$ is not tabulated in Ref.~\cite{formfac}. Following Refs.~\cite{formfac2, NRO}, it is approximated by that of the lighter 4$d$ element Y.

With these inputs, together with the overall scale factor from the nuclear refinement, the only free parameters are the basis vector coefficients $C_\nu$ of the selected irrep.

Magnetic Bragg intensities were collected at 4.25~T, near the inflection point of the field dependence (Fig.~\ref{fig:fieldscanop}(a)), and at 6~T, well above the transition. This enables a direct test of whether transverse and longitudinal staggered components coexist near the QCP, and whether the longitudinal component is suppressed at higher fields.

\begin{figure}[t]
\centering
\includegraphics[clip, trim=0.0cm 0.0cm 0.0cm 0.0cm ,width=\linewidth]{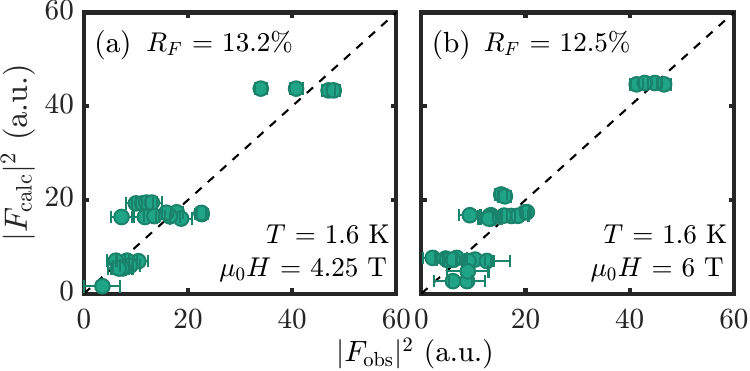}
\caption{\label{fig:refinefield} Calculated magnetic structure factors $|F_{\mathrm{calc}}|^2$ vs measured integrated intensities normalized by the Lorentz factor correction $\propto |F_{\mathrm{obs}}|^2$ at (a) 4.25~T and (b) 6~T. Dashed line indicates where the calculated and observed structure factor coincide.}
\end{figure}

\subsubsection{Vicinity of QCP (4.25~T)}

\begin{figure}[b]
\centering
\includegraphics[clip, trim=0.0cm 0.0cm 0.0cm 0.0cm ,width=\linewidth]{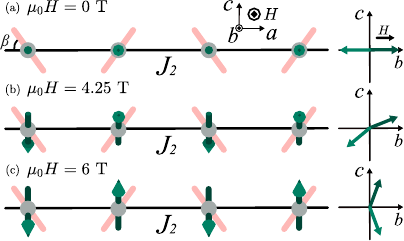}
\caption{\label{fig:magstruc} Schematic overview of the magnetic structures at $\mu_0H$ = (a) 0~T, (b) 4.25~T, and (c) 6~T. The dominant $J_2$ is projected as a chain. Rose lines indicate side views of single-ion anisotropy planes tilted with angle $\beta$ with respect to the \textbf{a} axis. Right inset shows the $bc$ projections of the two sublattices.}
\end{figure}

\begin{figure*}[t!]
\centering
\includegraphics[clip, trim=0.0cm 0.0cm 0.0cm 0.0cm ,width=\linewidth]{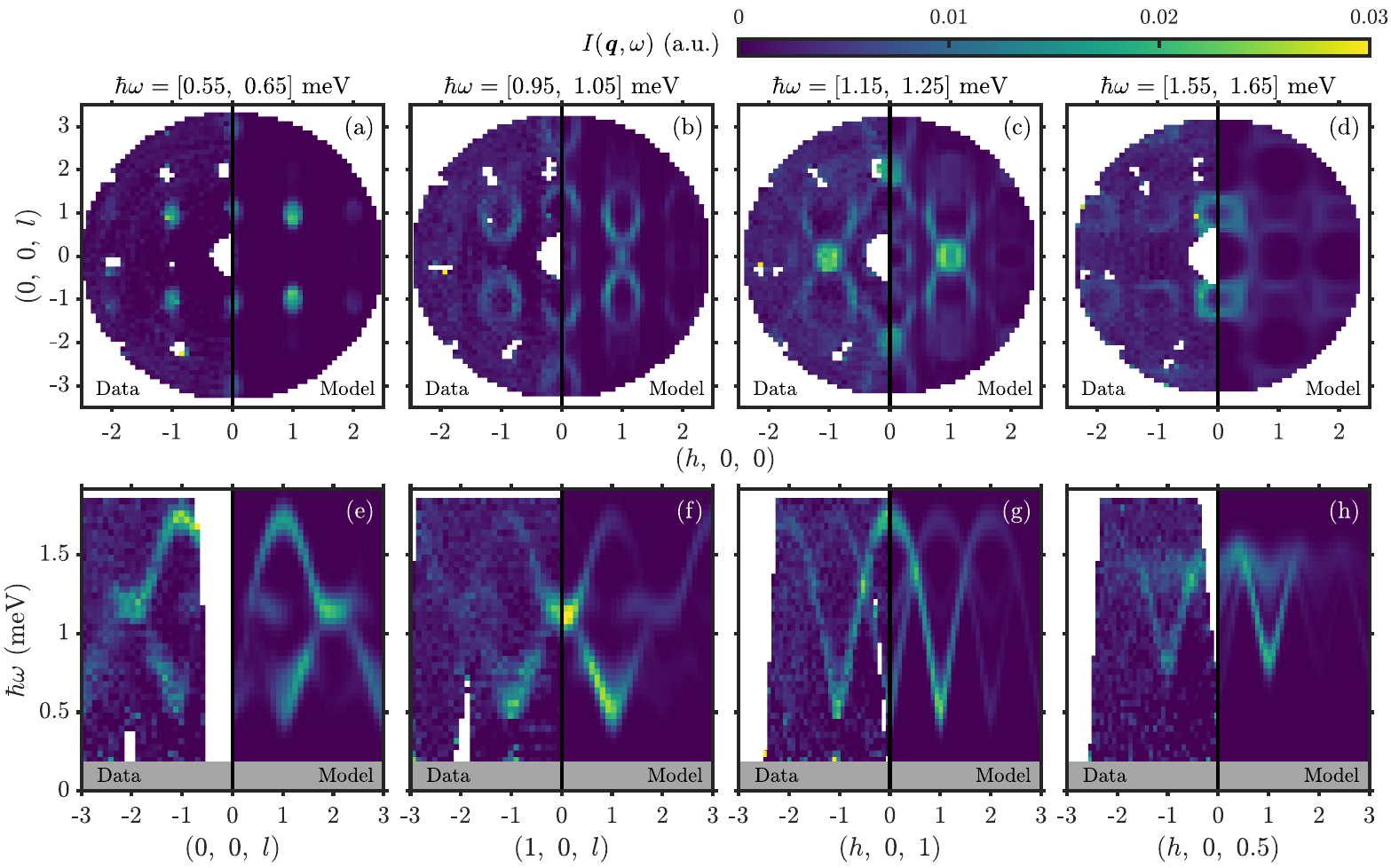}
\caption{\label{fig:INS} On the left side of each plot are false color representations of constant energy (a-d) and energy-momentum (e-h) projections of measured intensity in Cs$_2$RuO$_4$ at $T$ = 1.6~K. (a-d) The data are integrated in energy in the range $\pm$0.05~meV and binned in 0.06~\AA$^{-1}$. (e-f) The data are integrated perpendicular to the scan direction in the $k$-$l$ plane in the range $\pm0.1$~r.l.u. along $k$ and in the range $\pm0.1$~r.l.u. along $l$. On the right side of each plot are SU(3) spin-wave theory calculations based on the minimal parameters listed in Table \ref{tab:interactions}.}
\end{figure*}

The magnetic structure at 4.25~T was determined from 27 magnetic reflections collected as rocking curves with a counting time of 20~s/point. Integrated intensities were corrected for the Lorentz factor. Of all irreps compatible with the propagation vector, $\Gamma_2$ again provided the best fit. A good agreement with the data was achieved with two free parameters [$R_F$ = 13.2\%, Fig.~\ref{fig:refinefield}(a)]: $C_2$ = 1.43(7)~$\mu_{\mathrm{B}}$ and $C_3$ = 0.92(7)~$\mu_{\mathrm{B}}$, corresponding to the basis vectors $\psi_2$ (along \textbf{b}) and $\psi_3$ (along \textbf{c}), respectively.

An \textbf{a} component, variable magnetic domain mass ratios, and staggered \textbf{b} and \textbf{c} components on domains $B$ and $C$ were tested as additional parameters, but none improved $R_F$ by more than 2\% or significantly altered the solution for domain $A$ of interest, confirming the robustness of the minimal two-parameter solution.

The refined Fourier component of the staggered magnetic moment on Ru$_1$ is
\begin{equation}
    \textbf{\textit{A}}_1^{(0~0.5~0.5)} = (0,~1.43(7),~0.92(7))~\mu_{\mathrm{B}}.
\end{equation}
The uniform component along \textbf{b} $\textit{A}_{\textbf{b},1}^{(0~0~0)} = 0.14(3)~\mu_{\mathrm{B}}$ can be inferred from the bulk magnetization. The resulting magnetic structure including the accessible uniform \textbf{b} component, is shown schematically in Fig.~\ref{fig:magstruc}(b). At this field the longitudinal staggered component from the ZF ground state is retained, while a transverse component along \textbf{c} has emerged.

\subsubsection{Spin-flop phase (6~T)}

The magnetic structure at 6~T was determined from 30 magnetic reflections collected and corrected for the Lorentz factor. As at 4.25~T, $\Gamma_2$ provided the best fit among all compatible irreps. A minimal one-parameter solution using only refined coefficient $C_3$ = 1.73(2)~$\mu_{\mathrm{B}}$ [$R_F$ = 12.5\%, Fig.~\ref{fig:refinefield}(b)] was obtained. This is consistent with the full expected moment $g_\textbf{c}S$ = 1.7~$\mu_{\mathrm{B}}$ along \textbf{c}. The same robustness checks as at 4.25~T and an additional \textbf{b} staggered component confirmed that no additional parameters improve the solution by more than 2\%.

The refined Fourier component of the magnetic moment on Ru$_1$ is:
\begin{equation}
    \textbf{\textit{A}}_1^{(0~0.5~0.5)} = (0,~0,~1.73(2))~\mu_{\mathrm{B}}.
\end{equation}
Together with the uniform \textbf{b} component $\textit{A}_{\textbf{b},1}^{(0~0~0)} = 0.37(3)~\mu_{\mathrm{B}}$ inferred from the bulk magnetization, the resulting magnetic structure is shown in Fig.~\ref{fig:magstruc}(c). The longitudinal staggered component has vanished within resolution, leaving only a staggered moment along \textbf{c} and an observable uniform moment along \textbf{b}, consistent with a spin-flop state.

\subsection{\label{sec:INS}Neutron spectroscopy}

With the field evolution of the magnetic structure established, ZF neutron spectroscopy measurements were performed in the $(h~0~l)$ plane to further refine the model Hamiltonian.

\begin{figure*}[t]
\centering
\includegraphics[clip, trim=0.0cm 0.0cm 0.0cm 0.0cm ,width=\linewidth]{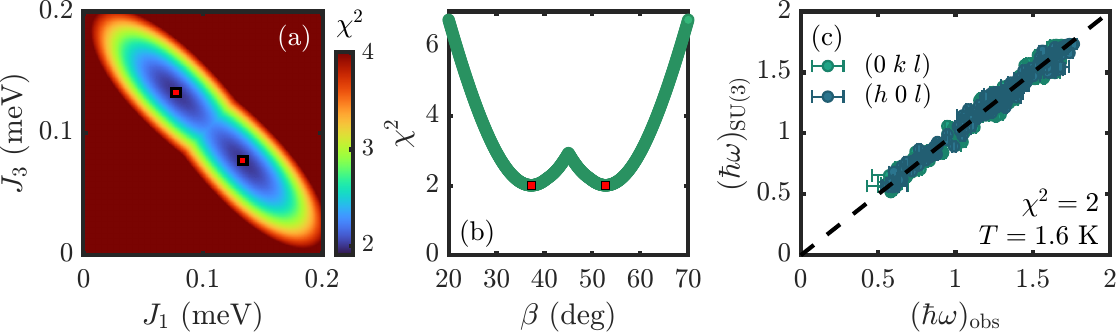}
\caption{\label{fig:INSLSQ} (a) False color plot of $\chi^2$ as a function of $J_1$ and $J_3$ and (b) $\chi^2$ vs the single ion anisotropy angle $\beta$, both while keeping other parameters constant. Red squares are minima $\chi^2$ solutions. (c) Calculated vs fitted energy positions from constant \textbf{\textit{q}}-scans. Calculated energies come from SU(3) spin-wave theory simulations.}
\end{figure*}

In Ref.~\cite{CRO}, domain mass ratios were used to deconvolute spectra measured in the $(0~k~l)$ plane and reconstruct the single-crystal response. This approach was possible because all three domains share the same scattering plane in that geometry, being related by rotations about \textbf{a}. In the present $(h~0~l)$ configuration, however, the two minority domains are oriented in the $(h~k~k)$ and $(h~-k~k)$ planes. As a result, the deconvolution procedure is no longer applicable. Instead, the contributions from all three domains are simulated simultaneously and combined using their respective mass ratios from the $(0~k~l)$ experiment for comparison with the data, as the same sample was used.

Representative constant-energy slices and energy–momentum projections measured in the $(h~0~l)$ plane are shown in Fig.~\ref{fig:INS}. Background subtraction was performed following the procedure of Ref.~\cite{CRO}. The data reveal that the complex spectrum of resolution-limited modes exhibits strong dispersion along $h$, confirming the three dimensional nature of the excitations.

The ZF spectrum was previously described using SU(3) generalized spin-wave theory (GSWT) with a minimal Heisenberg model including three of the first five nearest-neighbor interactions ($J_1$, $J_2$, $J_4$), a next-nearest-neighbor interaction $J_8$, and single-ion anisotropy parameters $D$ and $\beta$~\cite{CRO}. While this model reproduces the $(h~0~l)$ spectrum reasonably well, it fails to capture the curvature of the dispersion near $(0~0~1)$ along $h$ [Fig.~\ref{fig:INS}(g)]. This discrepancy originates from the near-indistinguishability of $J_1$ and $J_3$ in the $(0~k~l)$ geometry. Both interactions connect spins along \textbf{a} within the same zigzag chain motif, coupling ions with identical single-ion anisotropy plane orientations, and therefore enter the spin-wave Hamiltonian nearly equivalently if there is no sensitivity to the dispersion along \textbf{a}. Their relative magnitudes, however, strongly influence the dispersion curvature along $h$ near $(0~0~1)$, enabling their independent determination from the present dataset.

To resolve this, $J_3$ was included, and the resulting seven-parameter model was refined by fitting excitation energies extracted from 651 constant-$\bm{q}$ cuts across both geometries to the corresponding SU(3) GSWT calculations. Multiple least-squares fits were initialized from random starting points within physically reasonable bounds around the solution of Ref.~\cite{CRO} to explore the parameter landscape and identify local minima.

Two degeneracies were identified. First, $J_1$ and $J_3$ can be interchanged without affecting the fit quality, resulting in two equivalent minima, as illustrated in the $\chi^2$ color map in Fig.~\ref{fig:INSLSQ}(a). Second, the anisotropy angle $\beta$ exhibits a mirror degeneracy about $45^{\circ}$, corresponding to tilts of the easy axis toward either \textbf{a} or \textbf{c} (Fig.~\ref{fig:INSLSQ}(b)). Together, these yield four indistinguishable solutions. A unique solution was selected based on physical considerations: $\beta$ is chosen to tilt toward \textbf{c}, consistent with the observed high-field spin-flop structure featuring staggered moments along \textbf{c}, and $J_1 > J_3$, consistent with the shorter bond length. The resulting parameters are listed in Table~\ref{tab:interactions}, yielding an excellent fit with $\chi^2 = 2$. A comparison of observed and calculated excitation energies is shown in Fig.~\ref{fig:INSLSQ}(c).

\begin{table}[t]
    \centering
    \caption{Refined minimal model Hamiltonian parameters for Cs$_2$RuO$_4$.}
    \label{tab:interactions}
    \begin{tabular*}{\linewidth}{@{\extracolsep{\fill}}lccc}
        \toprule
         & Bond length (\AA) & Anisotropy & Value (meV) \\
        \midrule
        $J_1$ & 5.34 & $\parallel$     & 0.13(2) \\
        $J_2$ & 5.74 & $\perp$     & 0.44(3) \\
        $J_3$ & 5.95 & $\parallel$     & 0.08(5) \\
        $J_4$  & 6.44  & $\parallel$     & 0.22(2) \\
        $J_8$  & 8.62  & $\perp$      & -0.02(1)  \\
        $D$  & --  & -- & 0.26(5)  \\
        $\beta$  & --  & -- & 53(5)$^{\circ}$  \\
        \bottomrule
    \end{tabular*}
\end{table}

Due to the complex $\chi^2$ landscape, uncertainties derived from least-squares fitting alone are underestimated. Instead, uncertainties were estimated using a reverse Monte Carlo approach: 20~000 parameter sets were sampled around the selected solution and retained if $\Delta\chi^2 \leq 0.5$. The spread of accepted solutions defines the error bars reported in Table~\ref{tab:interactions}.

The table also includes the corresponding bond lengths and indicates whether each interaction couples spins with parallel or nearly orthogonal single-ion anisotropy planes. Simulated spectra, including contributions from all three crystallographic domains, are shown alongside the data in Fig.~\ref{fig:INS}. These have been convoluted with a Gaussian with a standard deviation based on the elastic resolution of the instrument. The model reproduces the key features of the spectrum with excellent agreement.

\section{\label{sec:DIS}Discussion}

The measured field evolution of the magnetic structure is in good agreement with predictions from both the toy chain and the full SU(3) model presented in Ref.~\cite{CRO}. The system evolves continuously from a N\'eel state in ZF, characterized by a longitudinal staggered moment along the effective easy \textbf{b} axis, through a regime near the QCP for which longitudinal and transverse staggered components coexist, to a high-field spin-flop phase in which only the transverse \textbf{c} component remains. This evolution is schematically illustrated in Fig.~\ref{fig:magstruc}. While a small staggered \textbf{a} component is predicted by the full SU(3) calculations, it is insensitive to the present refinements, however, its absence does not affect the overall behavior in the vicinity of the QCP.

The refined minimal Hamiltonian does not significantly modify the previously predicted field evolution of the magnetization, staggered magnetization, or excitation gap obtained from SU(3) simulations \cite{CRO}. However, the experimentally determined magnetic structure, specifically the identification of the transverse staggered component along \textbf{c} and a refined tilt angle $\beta > 45^\circ$, lifts the degeneracy of the high-field spin-flop state, favoring alignment along \textbf{c} over \textbf{a}.

To enable a more direct comparison with experiment, the staggered magnetization obtained from refined SU(3) calculations was used to simulate order-parameter field scans. Magnetic structure factors were calculated at successive fields and converted to peak intensities via the Lorentz factor, with a single overall scale factor applied to all reflections. The resulting curves, shown as red lines in Fig.~\ref{fig:fieldscanop}, exhibit good qualitative agreement with the experimental data.

The transition and saturation fields are overestimated in the SU(3) simulations. To account for this discrepancy, the simulated field axis was rescaled to match the experimental saturation field. This deviation likely reflects a renormalization of the exchange parameters in ZF. A definitive determination of the microscopic parameters would require inelastic neutron scattering in the fully polarized regime, but the necessary magnetic fields ($> 25$~T) remain challenging to access in neutron experiments.

An important open question is the direct spectroscopic observation of the gap closing at the QCP, which has so far only been inferred indirectly from thermodynamic measurements. High-resolution probes such as ESR or THz spectroscopy, would be ideally suited to track the field evolution of the excitation gap across the QCP.

\section{\label{sec:CON}Conclusion}

The $S=1$ compound Cs$_2$RuO$_4$ undergoes a spin-flop-like transition accompanied by a quantum critical point driven by strong frustration of the single-ion anisotropy. The QCP is associated with the spontaneous emergence of a transverse staggered magnetization component coexisting with the longitudinal one. Neutron diffraction measurements quantitatively confirm the predicted field evolution of the magnetic structure. Furthermore, the minimal model Hamiltonian is refined based on the combined experimental results.

\section*{\label{sec:ack}Acknowledgements}
This work is supported by a MINT grant of the Swiss National Science Foundation. This work is based on experiments performed at the Swiss spallation neutron source SINQ, Paul Scherrer Institute, Villigen, Switzerland PSI (CAMEA ID: 20241402, Zebra ID: 20240902). We acknowledge the beam time allocation at Institut Laue Langevin, Grenoble, France (D23 ID: CRG-3277).

\bibliography{bibliography.bib}

\end{document}